\documentclass[preprint]{aastex631}
\definecolor{mygreen}{RGB}{0, 185, 118}

\usepackage{amsmath, amssymb, amsfonts, latexsym}        

\shorttitle{Influence of the magnetic environment in deflections of solar eruptions}
\begin{document}

\title{Analysis of solar eruptions deflecting in the low corona: influence of the magnetic environment}

\correspondingauthor{Abril Sahade}
\email{asahade@unc.edu.ar}

\author[0000-0001-5400-2800]{Abril Sahade}
\affiliation{Heliophysics Science Division, NASA Goddard Space Flight Center, Greenbelt, MD, 20771, USA.}

\author[0000-0002-8164-5948]{Angelos Vourlidas}
\affiliation{The Johns Hopkins University Applied Physics Laboratory, Laurel MD 20723, USA.} 

\author[0000-0000-0000-0000]{Cecilia Mac Cormack}
\affiliation{Heliophysics Science Division, NASA Goddard Space Flight Center, Greenbelt, MD, 20771, USA.} 
\affiliation{The Catholic University of America, Washington, DC 20064, USA}

\begin{abstract}
    Coronal mass ejections (CMEs) can exhibit non-radial evolution. The background magnetic field is considered the main driver for the trajectory deviation relative to the source region. 
    The influence of the magnetic environment has been largely attributed to the gradient of the magnetic pressure. In this work, we propose a new approach to investigate the role of topology on CME deflection and to quantify and compare the action between the magnetic field gradient (`gradient' path) and the topology (`topological' path).

    We investigate 8 events simultaneously observed from Solar Orbiter, STEREO-A and SDO; and, with a new tracking technique, we reconstruct the 3D evolution of the eruptions. Then, we compare their propagation with the predictions from the two magnetic drivers. We find that the `topological' path describes the CME actual trajectory much better than the more traditional `gradient path'.  
    
    Our results strongly indicate that the ambient topology may be the dominant driver for deflections in the low corona, and that presents a promising method to estimate the direction of propagation of CMEs early in their evolution.
\end{abstract}

\section{Introduction}

Coronal mass ejections (CMEs) are the drivers of the strongest geomagnetic storms and a major concern in space weather. They are manifestations of the ejection of a magnetic flux rope (MFR) that connects the CME and its early evolutionary processes (e.g. prominence/filament eruptions, flares, and cavity formation and evolution) with the source region in the lower corona \citep[e.g.,][]{Webb2012,Patsourakos2020}. The reliable prediction of the eruption and its subsequent trajectory is crucial to assessing its potential geoeffectiveness. 

Early on it was noted that some CMEs undergo equatorward deflection from the radial direction in the low corona \citep{macqueen1986}. This was interpreted as evidence of the influence of the background magnetic field irrespective of whether the event originated in a quiescent region or was accompanied by a flare. Subsequent observations from LASCO \citep[e.g.][and references therein]{Kilpua2009} demonstrated that the equatorward deflection is more prominent in events originating from polar crown filament eruptions at high latitudes. But the single-viewpoint observations could not (and still cannot) accurately determine the direction of propagation of an event.

The multipoint observations enabled by the launch of the \textit{Solar TErrestrial RElations Observatory} \citep[STEREO,][]{STEREO_2008} twin spacecraft (STA and STB, hereafter), along with the development of various reconstruction tools \citep[e.g.,][]{mierla2008,maloney2009,temmer2009,Thernisien2009,Kwon2014,Zhang2021A&A} allowed the determination of the CME source regions and the computation of the three-dimensional (3D) path of CMEs.

Ignoring deviations driven by CME-CME interactions (relevant mainly in the outer corona and beyond) or magnetic imbalances within the MFR, the 3D analysis demonstrated that low corona deflections are mainly controlled by the background magnetic field \citep[e.g.,][]{Bemporad2012,Kay2013,kay2015,Liewer2015,Sieyra2020}. Lorentz forces are generally dominant in the low corona (within 3 solar radii or so). Therefore, most of a magnetic-driven deflection is expected to occur during the initial stages of a CME relatively close to the Sun.

\cite{Shen2011} proposed a model to quantify the deflection in terms of the magnetic pressure gradient. \cite{gui2011} extended the study, derived the trajectories of 11 CMEs above $2.3\,R_\sun$, and found a positive correlation between the deflection rate and the strength of the gradient. The main conclusion from these works is that CMEs tend to deflect toward the region of low magnetic energy density. This result is in good agreement with the equatorward deviations of CMEs during low solar activity \citep{cremades2004}, and more recent studies showing that CMEs preferentially deflect from their source region to the heliospheric current sheet (HCS) or neighboring pseudostreamers \citep[e.g.][]{Zuccarello2012,Wang2020JGRA,Wang2023AdSpR}. Additionally, the magnetic pressure gradient produced by strong active-region fields has been referred to as the cause of non-radial propagation for active-region CMEs \citep[e.g.,][]{Liewer2015,kay2015}. 

However, an exhaustive analysis carried out by \cite{Sieyra2020} showed that the deflection of the prominence and CME observed at low heights does not tend to align with the magnetic pressure gradient. One problem with the gradient interpretation is that structures such as coronal holes (CHs) exert minimal changes on the gradient. Consequently, even though CHs have a significant impact on the evolution of CMEs \citep[e.g.,][]{Cremades2006,Kilpua2009,Gopalswamy2009,cecere2020,Sahade2020}, they are ``invisible'' to the magnetic pressure gradient.

An alternative factor to consider is the presence of magnetic null points. These regions of low-to-null magnetic pressure have been shown to influence the direction of propagation of MFRs \citep[e.g.,][]{Panasenco2013,Sahade2021, Sahade2022, Sahade2023}. The deflections are also attributed to the topology of the ambient magnetic field, indicating the ability of the field lines to ``channel'' the MFR. 
In this sense, magnetic null points can be understood not only as potential wells from an energetic perspective but also as flux separators from a topological perspective. In \cite{Sahade2023}, we highlight the key role of null points in the early evolution of erupting MFRs and suggest that the determination of the ambient magnetic topology should significantly improve our understanding of the early trajectories of CMEs. The question then arises on which of the two deflection agents determines the early CME path; the magnetic field gradient or the null point(s) location together with the field line orientation.

We investigate this question here via a detailed analysis of CMEs observed from multiple viewpoints to determine their 3D evolution in the low corona. We use the Potential Field Source Surface model \citep[PFSS,][]{PFSS_2003SoPh} for the coronal magnetic field reconstruction and the definition of two magnetic paths. We form a simple hypothesis; namely, if an eruption deviates solely due to the magnetic field gradient, it should follow the `gradient' path, but if the deviation is driven by the magnetic topology, the event should follow the `topological' path.
 
The remainder of the manuscript is organized as follows. In Section~\ref{sec:data}, we describe the data and methodology used to compute: (1)  the actual eruptive path of the CME; (2) the `gradient' path defined by the minimal magnetic pressure gradient; and (3) the `topological' path defined by the overlying magnetic field. The results are presented in
Section~\ref{sec:results} along with comparisons amongst the three paths, including an error analysis. We conclude in Section~\ref{sec:conclusion}.

\section{Observations and methodology} \label{sec:data}
\subsection{Events selection}

\begin{deluxetable}{c | l r r r}
\tablecaption{Event list and coordinates.\label{tbl-events}}

\tablehead{
\colhead{Event date}  & $(\phi_0,\theta_0)$ & $\Delta\phi_M$ & $\Delta\theta_M$ & $\psi$\\
\colhead{\small{[YYYY-MM-DD]}}& \colhead{[ ($^\circ$,$^\circ$) ]}& \colhead{[ $^\circ$ ]}& \colhead{[ $^\circ$ ]} &\colhead{[ $^\circ$ ]}
}

\startdata
  2021-02-22 & $(272,-36)$ & $6$& $-2$ & $5$ \\
  2021-03-21 & $(264,-32)$ & $9$& $2$ & $7$ \\
  2021-04-26 & $(210,-52)$ & $-6$& $26$ & $28$ \\
  2021-12-26 & $(269,-35)$ & $10$& $8$ & $9$ \\
  2021-12-31 & $(248,-52)$ & $-12$& $17$ & $18$  \\
  2022-03-19 & $(170,-37)$ & $-5$& $-2$ & $3$  \\
  2022-03-20 & $(161,\quad\:7)$ & $-3$& $8$ & $7$  \\
  2022-03-28 & $(\,\,\,87,\,\,\,\,14)$ & $8$& $-6$ & $4$  
\enddata
\tablecomments{The first two columns display the event date and initial position in Carrington coordinates. Third and fourth columns display the maximum deflection in longitude and latitude, respectively. Positive values indicate deflections to the west and north; and negative values indicate deflections to the east and south, respectively. The 3D deflection angle $\psi$ at $2.5\,R_\sun$ is given at the last column.}
\end{deluxetable}

Because we want to assess the ability of magnetic factors to predict the propagation direction of eruptive events in the low corona, we select events that meet the following criteria.
First, to have the most accurate magnetic field information in the source region, we select events on the visible disk from the \textit{Solar Dynamics Observatory} \citep[SDO,][]{SDO_2012SoPh}. The PFSS model assimilates Helioseismic and Magnetic Imager \citep[SDO/HMI,][]{HMI2012} magnetograms, updating the central $120^\circ$ every six hours. 
Secondly, we choose events observed from three viewpoints to ensure a robust tracking of the eruption. We use the data provided by the \textit{Atmospheric Imaging Assembly} \citep[SDO/AIA,][]{AIA_2012SoPh}, the wavelet-enhanced images \citep{Stenborg2008} from \textit{Extreme-Ultraviolet Imager }\citep[STA/EUVI,][]{SECCHI_2008}, and the Solar Orbiter \textit{Extreme Ultraviolet Imager} \citep[SolO/EUI,][]{EUI_2020A&A} to follow the prominence material and, when visible, the associated CME in the low corona. Also, to better model the CME in the coronagraph images we require an angular separation greater than $30^\circ$ between STA and the \textit{Large Angle and Spectrometric Coronagraph Experiment} \citep[SOHO/LASCO,][]{LASCO_1995SoPh}. That constrained the observations to the period from February 2021 to May 2022. It is worth mentioning that SolO/EUI data are missing from May to December 2021. 

\begin{figure*}[ht]
    \centering
    \includegraphics[width=0.8\textwidth]{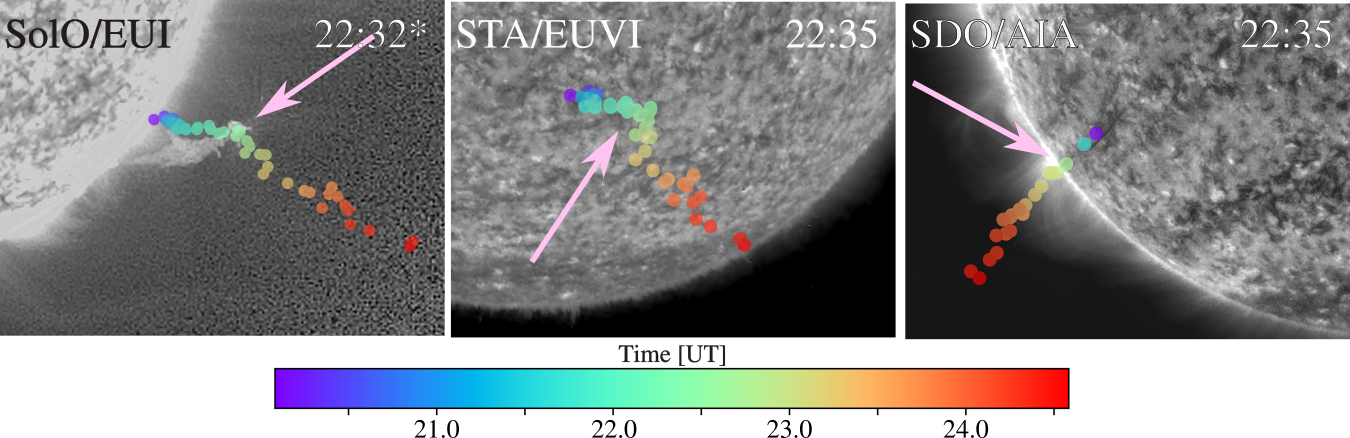} \\
    \hspace{1ex} \\
    \includegraphics[width=0.8\textwidth]{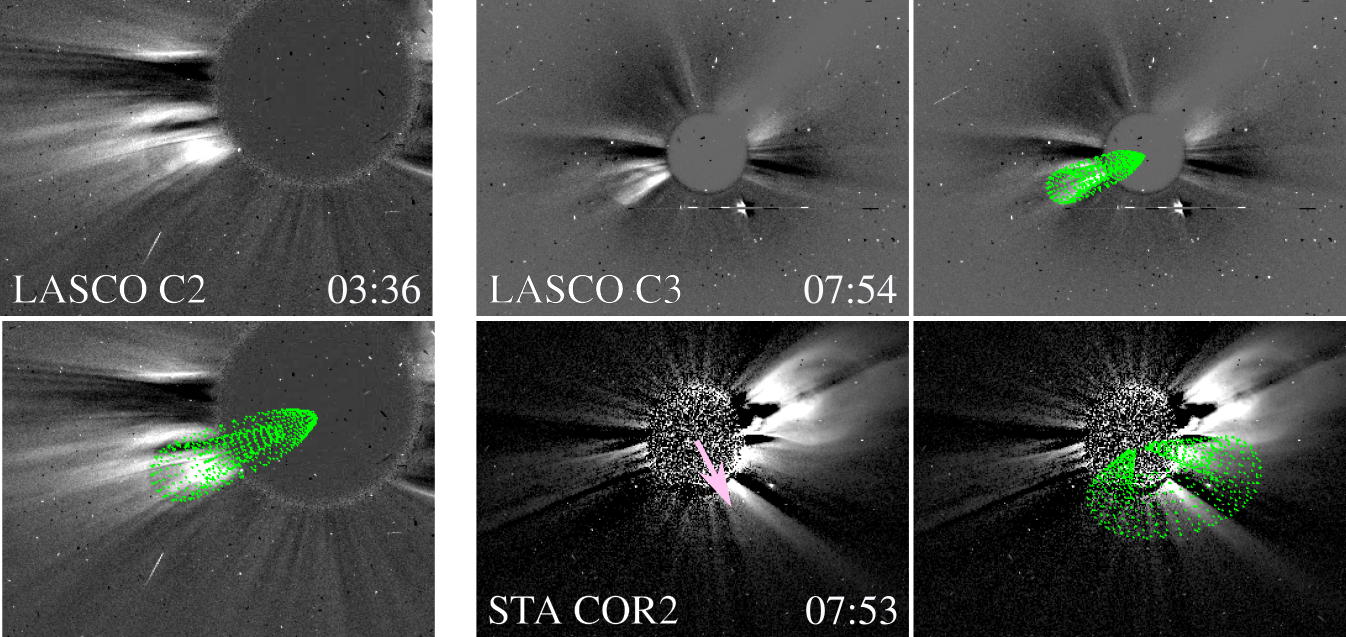}
    \caption{\textit{Upper panels:} Triangulated coordinates for the 2021-03-21 filament eruption. Color dots represent the mean measure of the coordinates for each time, as indicated by the colorbar. Each panel shows the filament as observed by SolO, STA, and SDO at 22:35 (22:32 for SolO is the time at 1 AU). The pink arrows indicate the filament apex at the time of the observation.\textit{Lower panels:} Non-radial GCS applied to the 2021-03-21 CME, observed on the 2021-03-22 from STA and LASCO. The left panels show the CME early observed from the single view point of LASCO C2 and the GCS mesh. The middle panels show the CME observed simultanously from LASCO C3 and STA COR2 at 7:54 UT, and the corrresponding GCS mesh (right panels). }
    \label{fig:tp}
\end{figure*}

The eventual event sample consists of 8 filament-related CMEs. We discarded flare events since it was impossible to reconstruct their trajectory with good spatial resolution (given the cadence of the instruments), and  filament eruptions without a distinguishable CME from both SOHO/LASCO and STA/COR1 or COR2. The sample includes from radially propagating to highly deflected events (by more than $20^\circ$ below $2.5\,R_\sun$).
Table~\ref{tbl-events} lists the event dates, their source region Carrington longitude and latitude $(\phi_0,\theta_0)$, the maximum longitudinal and latitudinal deflection ($\Delta\phi_M$ and $\Delta\theta_M$, respectively). Some events presented a growing trend in their deflection, while others reversed their motion in longitude, latitude or both directions. In the case of the former, $\Delta\phi_M$ and $\Delta\theta_M$ correspond to the deflection at $2.5\,R_\sun$. Negative values in $\Delta\phi_M$ indicate deflections to the east, and negative values in $\Delta\theta_M$ indicate deflections to the south. Conversely, positive values of $\Delta\phi_M$ and $\Delta\theta_M$ indicate deflections to the west and north, respectively. The last column of Table~\ref{tbl-events} shows the 3D angle of deflection $\psi$ at $2.5\,R_\sun$, given by the spherical angle:

\begin{equation*}
\psi(h)=\arccos [\sin (\theta_0) \sin (\theta_h) + \cos (\theta_0)  \cos (\theta_h)  \cos(\phi_0-\phi_h)],
\end{equation*}
\noindent
where $\phi_h$ and $\theta_h$ are the Carrington longitude and latitude at a given height $h$. Note that the angle $\psi$ varies uniformly in the sphere, independently of the latitudinal location of the eruption, giving a better notion of the magnitude of the deflection and combining the non-radial displacements in both latitude and longitude. We choose the 2021-03-21 event as an example to illustrate our methodology. With the aim of provide reproducible results and available for other possible researches, all our data and pipelines are available at \url{https://github.com/asahade/Events}. 

\subsection{3D coordinates tracking and eruptive path}\label{ss:eruptiveP}

To reconstruct the 3D trajectory of each filament and its different segments we apply the tie-pointing technique \citep[see, e.g.,][]{Inhester2006}. This technique uses observations of the same coronal feature from two telescopes at different vantage points to estimate its 3D location within the telescopes field of view (FOV). 

We use here, for the first time, the  \texttt{scc$\_$measure3.pro} \citep{scc_measure3} routine, an extension made by the lead author (A. Sahade) to extend the main triangulation routine (\texttt{scc$\_$measure.pro})  to include three viewpoints (available at \textit{SolarSoft} and \url{https://github.com/asahade/IDL}). The addition of a third viewpoint in the computation of the trajectory is important to reduce the uncertainties present when the FOVs of two telescopes are too similar or when the tracked feature is extended along the epipolar plane. An example is provided in the Appendix.

To reduce uncertainties in the triangulation of filament positions, we required observations made as close in time as possible. Because of the elliptical orbit of SolO, its distance from the Sun varies significantly (ranging from approximately 0.3 AU to over 1 AU). Consequently, the observation times need to be corrected for the difference in the light propagation time between SolO and SDO or STA. Therefore, we use the header flag \textit{date\_ear}, which provides the light arrival time at 1 AU, to pair the images accurately.

\begin{figure}
    \centering
    \includegraphics[width=0.47\textwidth]{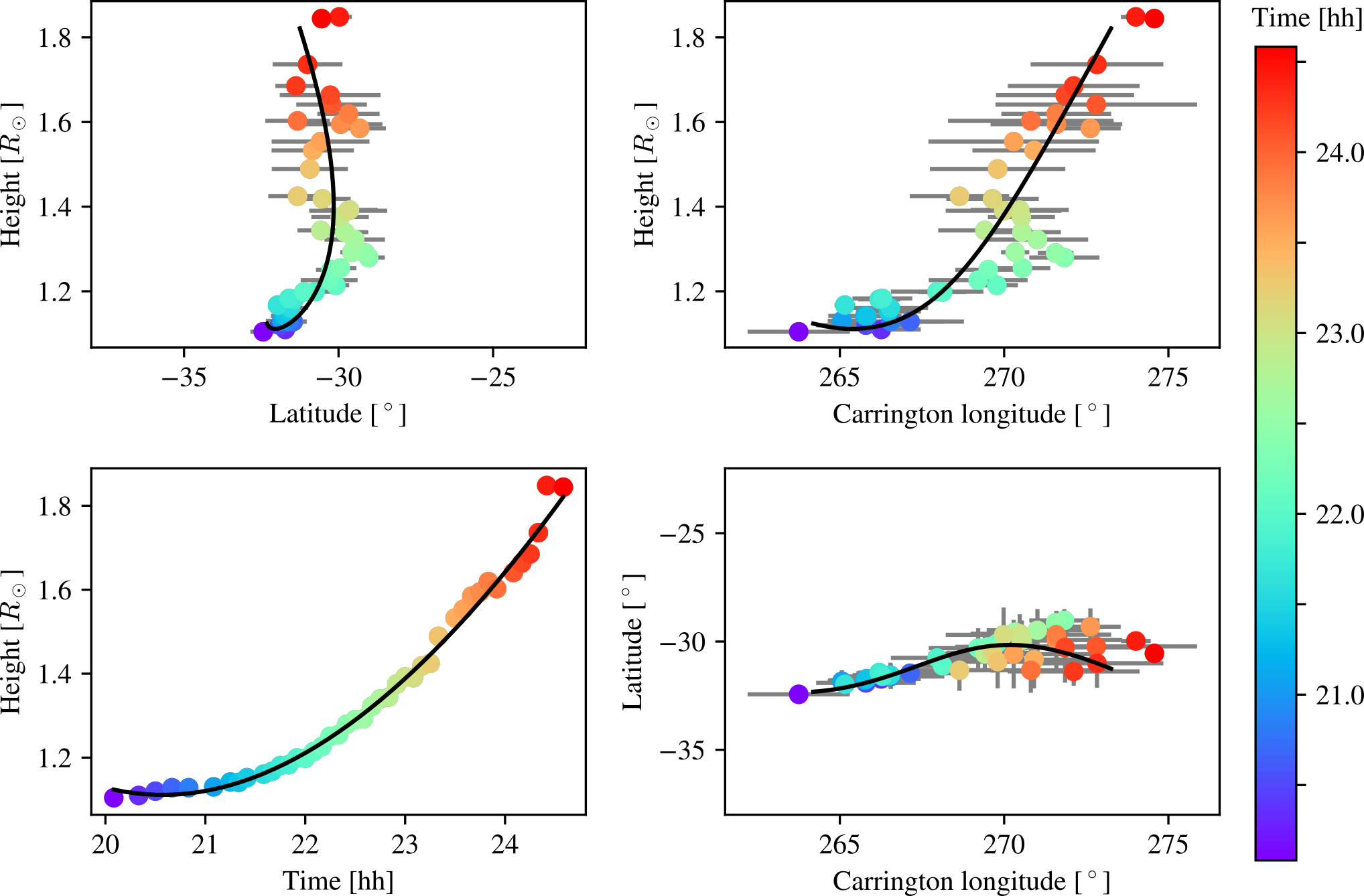}
    \caption{Triangulated and fitted coordinates for the 2021-03-21 filament eruption. Color dots idem Figure~\ref{fig:tp}. The grey bars indicate the standard deviation at each time (for the radial direction they are smaller than the dots size). Black lines are the fitted coordinates converted from the cartesian splines.}
    \label{fig:track_fit}
\end{figure}

\begin{figure*}
    \centering
    \includegraphics[width=0.96\textwidth]{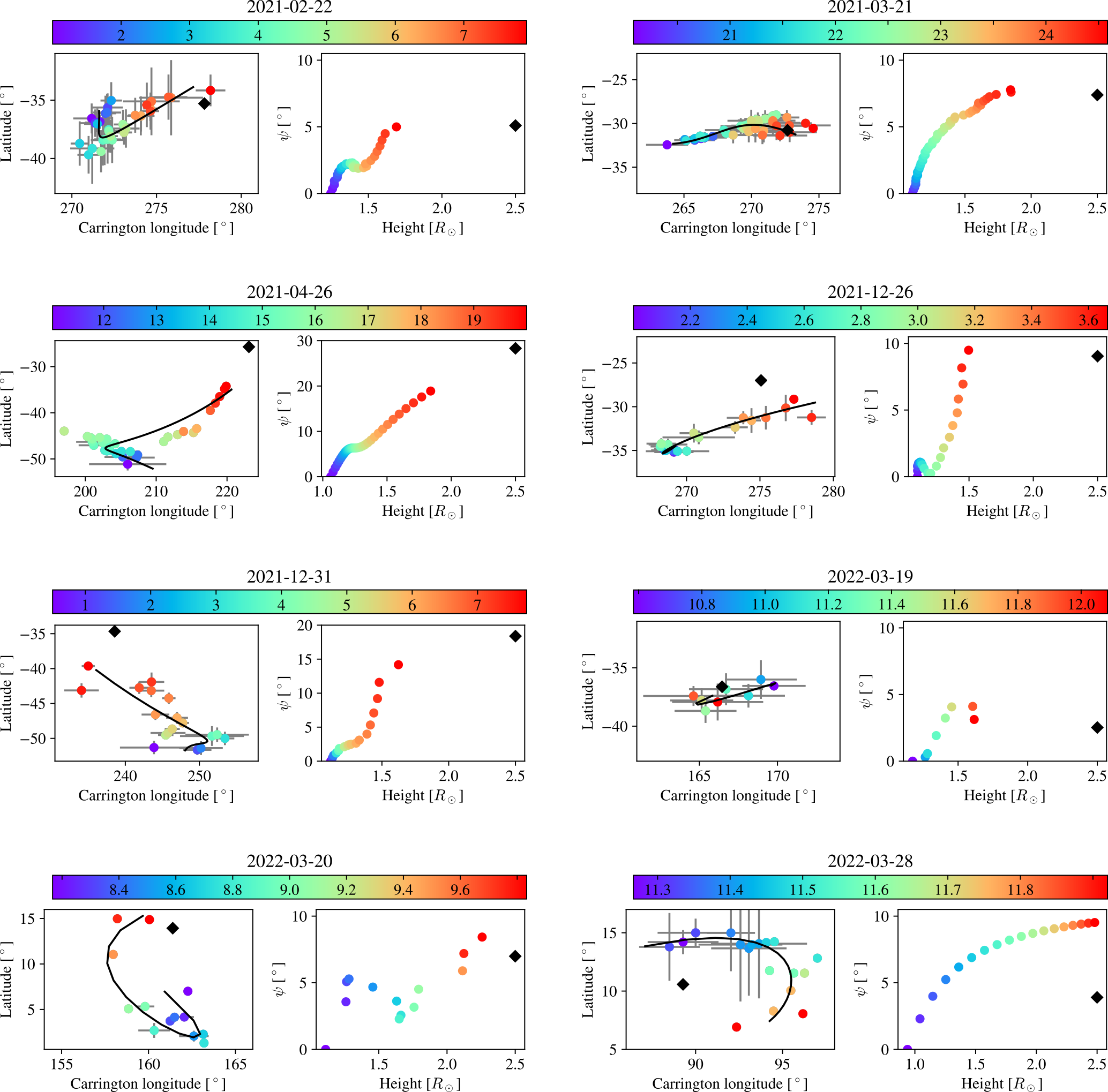}
    \caption{Projected deflection for all events. For each event, the left sub-panel shows the triangulated (colored dots scaled by time [UT]) and fitted curve (black line) for the latitude and Carrington longitude. The black diamond shows the interpolated CME position at $2.5\,R_\sun$. Right sub-panel of each date shows the 3D angle of deflection $\psi$ in function of height.}
    \label{fig:all-events}
\end{figure*}

To track the filament structure, we select and track different points around its apex. The distribution of these points can be more widespread when the apex position is not clearly defined. As SolO/EUI FSI304 provides observations every 10 minutes or longer, we track the filament structure with that cadence. Upper panles of Figure~\ref{fig:tp} shows the mean position of the triangulated features at each time, from the point of view of SolO, STA, and SDO, for the 2021-03-21 event. The images are a combination of 304 {\AA} and 171 {\AA} filters to enhance the filament, as observed at 22:35 UT (22:32 for SolO). The pink arrows indicate the triangulated dot (light green) corresponding to the observed time, and the color bar displays the time code of the dots. In the right panel (SDO image) the triangulated dots impeded the visibility of the filament, so fewer dots are displayed in the panel. From the 3D Cartesian coordinates, we modeled the prominence path as the cubic spline as a function of time for all the triangulated points (using the function \texttt{scipy.interpolate.UnivariateSpline}). We choose to fit the trajectory in Cartesian coordinates since the changes in the three coordinates are equivalent ($dx=dy=dz$). For clarity, here we present the results in spherical coordinates. Figure~\ref{fig:track_fit} shows the 3D spherical coordinates for the 2021-03-21 event. The upper panels show the evolution of the filament apex in Carrington latitude 
(left) and  longitude (right). In both cases, the vertical axis indicates the height of the filament to simplify the interpretation. Colored dots display the median coordinate of the apex at each time and its standard deviation (grey bars). Time is colorcoded  as shown in the colorbar. The bottom-right panel displays the overall deflection of the filament from its initial position (violet dot) to the last measurement (red dot). The black lines correspond to the fitted splines transformed to spherical coordinates. The error associated with the prominence coordinates is given by the standard deviation of each measurement unless it is smaller than the error of the fitting.

We also fitted the Graduated Cylindrical Shell \citep[GCS,][]{Thernisien2006,Thernisien2009} model to estimate the trajectory of the ejected material in the coronagraph observations (and EUV observations when possible). To reproduce the deflection of the CME in the low corona, we use the non-radial GCS model, which allows us to fix the CME footpoints in the solar surface while the CME front can vary in both latitude and longitude. This tool is available in \textit{SolarSoft} running the routine \texttt{rtcloudwidget}. In this study, we reconstruct the CME trajectories by adjusting the following free parameters: latitude, longitude, tilt angle, height, half angle, aspect ratio, and non-radial tilt. The latter parameter enables the modification of the angle subtended by the CME axis and the radial plane defined by the footpoints and the solar center. This allows us to better capture the CME front while CME footpoints remain within the source region. Lower panels of Figure~\ref{fig:tp} shows the CME of our example event. Due to the source region position it produced a very faint CME from STA point of view. Using the source region position and the single view-point reconstruction from LASCO C2 (left panel of Figure~\ref{fig:tp}) we constrained the parameters to obtain  more reliable parameters. The middle and right panels of Figure~\ref{fig:tp} shows the CME at 2021-03-22 7:54 UT, when observed simultaneously from LASCO C3 ans STA COR2. When possible (data available and simultaneous observation), we included the third viewpoint provided by the Solar Orbiter Heliospheric Imager \citep[SoloHI][]{SOLOHI2020}, to reduce the uncertainties on the GCS reconstructions. 

To model the CME path, we perform a linear fitting on the CME trajectory using the last tracked point of the prominence and the coordinates of the GCS cross-section center. The associated error for these coordinates is defined by the radius of the GCS cross-section, because our calculations are focused on determining the prominence apex, which is expected to be located within that segment of the MFR. 

In this work, our aim is to compare the observationally-derived 3D trajectories of the events with the paths derived from the PFSS model (with the source surface at $r_\text{ss}=2.5\,R_\sun$). We define the eruptive path of an event as the modeled prominence path (cubic spline), with the addition of the modeled cross-section center coordinates of the GCS at $r_\text{ss}$. 
Figure~\ref{fig:all-events} shows the deflection estimates for all the events analyzed in a lat-long map (left subpanels) and shows the evolution of the 3D angle of deflection, $\psi$, with the height of the eruption (right subpanels). The colored dots indicate the time (UT) per color bar. The black diamonds in the right sub-panels mark the angle $\psi$ at $2.5\,R_\sun$. 

The figure provides a comparative summary of our 3D fits. Most of the prominences appear to evolve along a straight, but non-radial, path away from the source region above $1.3\,R_\sun$. Two events (2022-03-20 and 2022-03-28) exhibit unusual curved paths. Both events presented a high rotation of the filament, as observed from its 3D reconstruction, and their onsets were associated with flares, which also made them fast eruptions. 

Two events (2021-04-26 and 2021-12-31) exhibit a high 3D deflection ($\psi>15^\circ$), two events (2021-02-22 and 2022-03-19) exhibit a mostly radial path ($\psi\leq 5^\circ$), and the rest of the events showed a moderate deflection ($\psi\leq 10^\circ$) below $2.5\,R_\sun$. Generally, $\psi$ does not evolve linearly with height. This indicates that the prominence is not simply ejected along a non-radial path, due to, say, the reconnection evolution but the deflection process continues to operate throughout the low corona. We will return to these observations later in the discussion.

\subsection{The gradient path}

\begin{figure}
    \centering
    \includegraphics[width=0.40\textwidth]{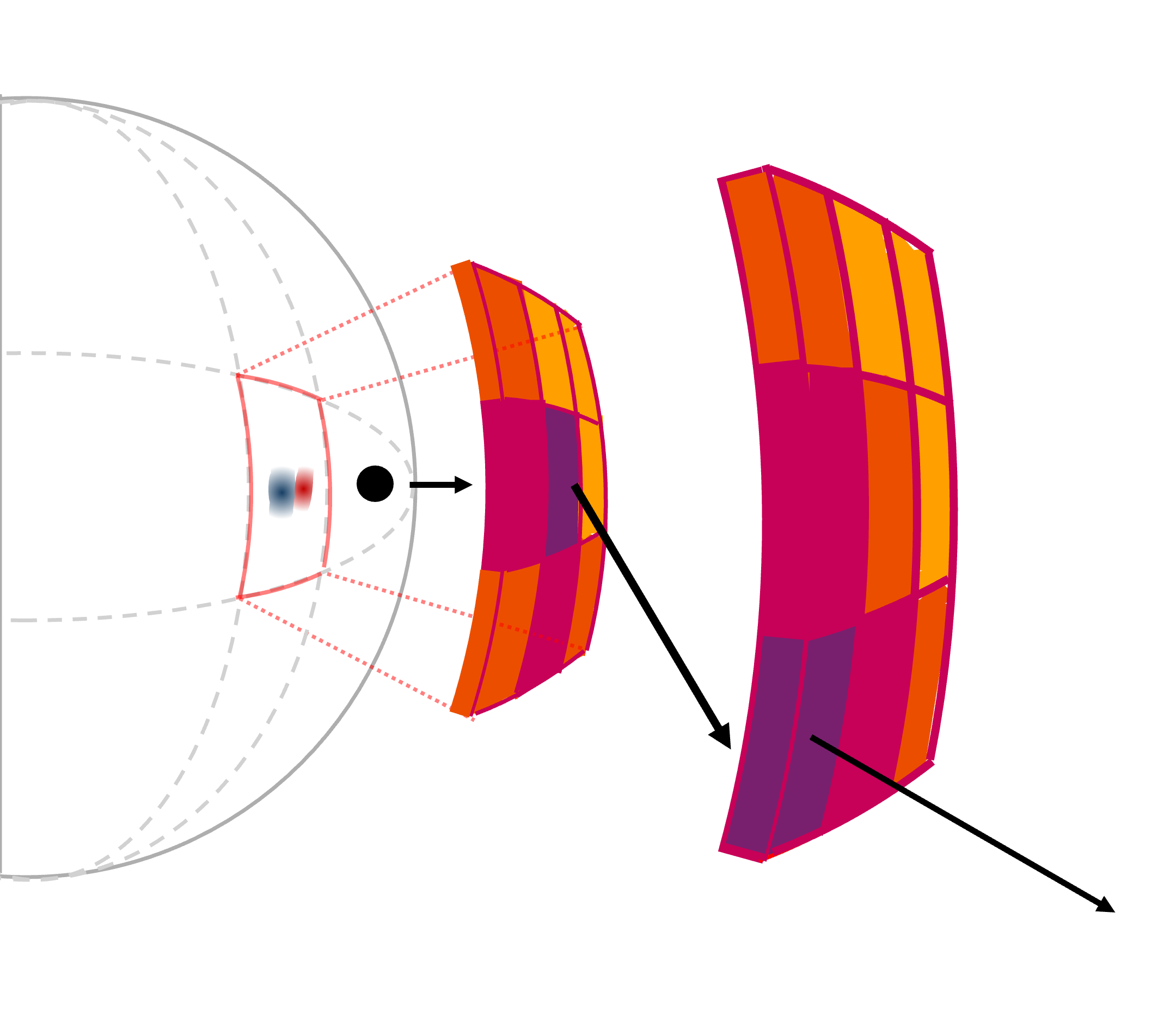} \\
    \includegraphics[width=0.30\textwidth]{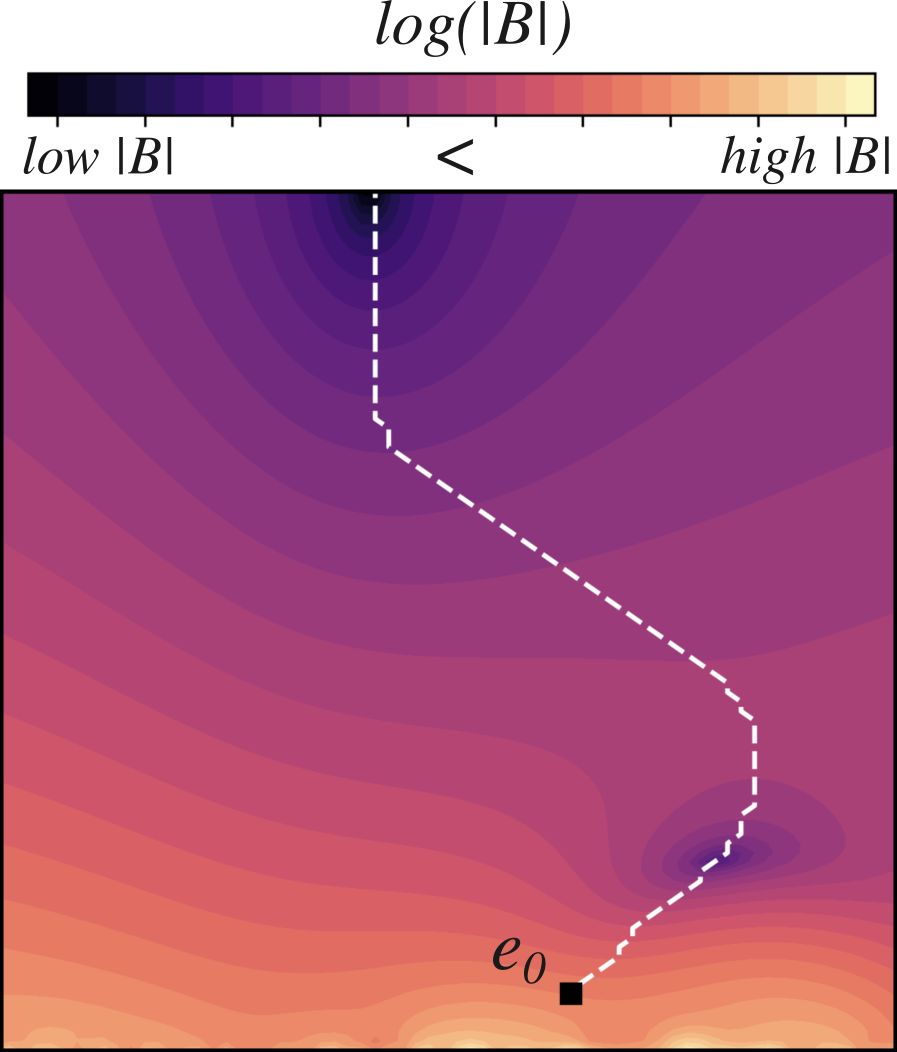}
    \caption{Upper panel: Scheme of the voxel selection for the `gradient' path. From an initial position $e_0$ (black dot) the nine upper voxels are considered and the one with minimal magnetic energy (darker one) is selected. From that position the next nine upper voxels are considered to trace the `gradient' path. Lower panel: For simplicity, an example of the gradient path in a bi-dimensional magnetic field $B$ is given. The white dashed line indicates the `gradient' path from the initial position $e_0$ (black square). The color scale indicates lower magnetic field magnitude when darker (and purpule), and higher $|B|$ when lighter (and yellow).}
    \label{fig:grad-sketch}
\end{figure}

As mentioned in the previous Section, the magnetic pressure gradient (with the conventional minus sign in front, i.e., $\vec{G}=-\nabla \frac{B^2}{2\mu_0}$) seems to influence the deflection of solar eruptions by guiding them to low magnetic energy regions such as null points, pseudostreamer spines, HCS, or simply away from active regions. In this work, we reexamine the ability of the ``gradient path'' i.e. the path toward the minimal magnetic energy region to account for the eruption trajectory. Instead of computing the change of the magnitude and direction of the gradient at each step of the eruption evolution, we derive the minimal gradient path from the source region. Starting from the initial position, $e_0$, of the eruptive path (at the source region), we select the next point of the path as it has the lowest magnetic energy ($B^2$). At each iteration, the radial direction increases by one grid point and the nine neighboring voxels (including diagonal voxels) above the actual position are considered for the selection. Figure~\ref{fig:grad-sketch} upper panel shows a diagram of the 3D voxel selection. The lower panel of Figure~\ref{fig:grad-sketch} shows a 2D example of the resulting gradient path (white dashed line) from the initial position $e_0$ (black square). In the example, the magnetic field has a low null point to the right of the initial position, and a global minimum, at the HCS location, to the left; the gradient path deviates towards both of them.

The coronal magnetic field is computed with the PFSS model. We select SDO/HMI magnetograms hours before the eruption take place as input for the PFSS model. We recompute the magnetic fields from the potential coefficients to obtain a radial resolution of $\delta r= 0.016\,R_\sun= 11\,\text{Mm}$, and angular resolution $\delta \psi= 0.94^\circ$. Nevertheless, the method to calculate the `gradient' path is independent of this specific selections and can be applied to any magnetic field reconstruction.

\subsection{The topological path}

\begin{figure}
    \centering
    \includegraphics[width=0.48\textwidth]{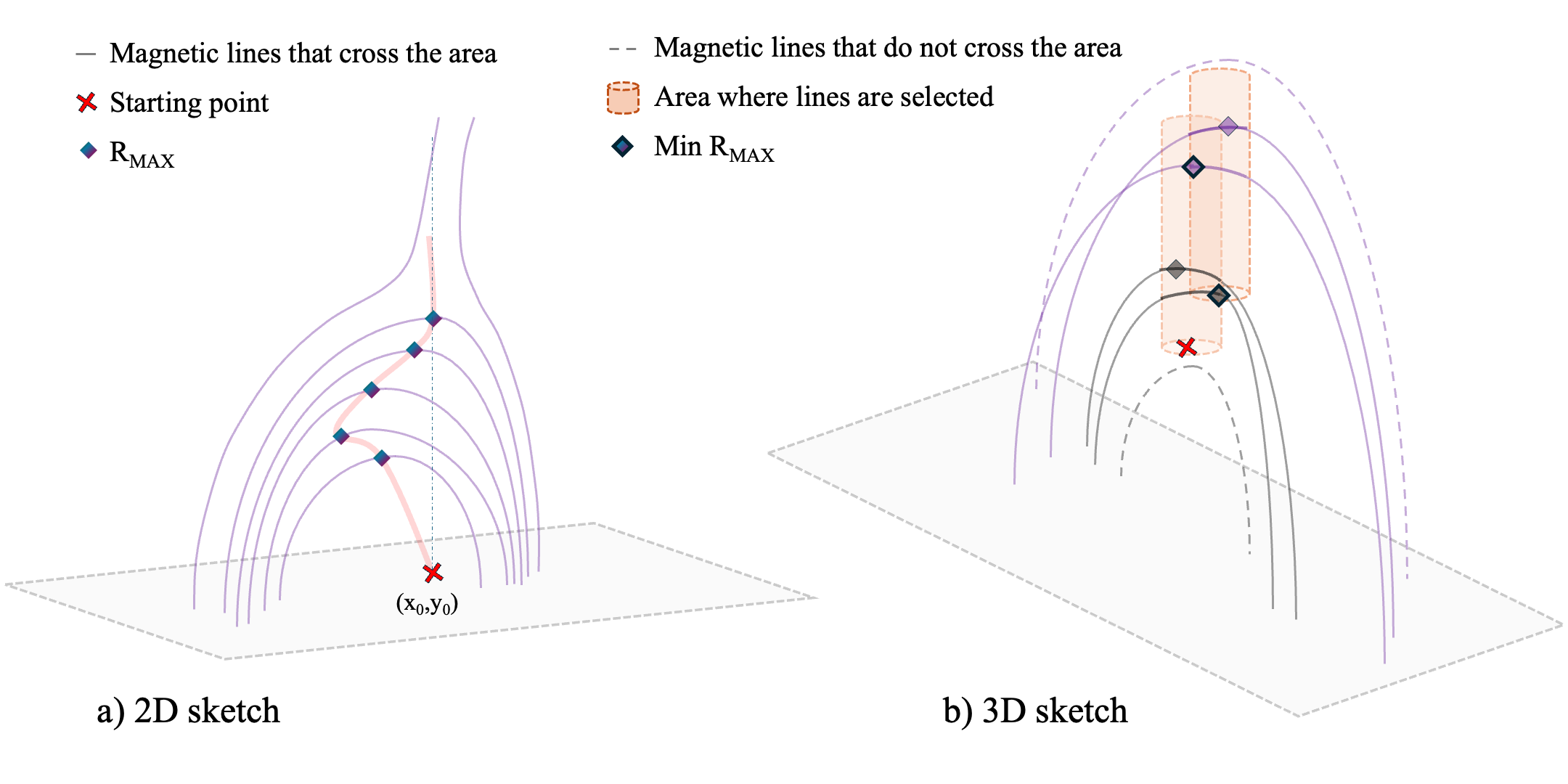} \\
    \includegraphics[width=0.35\textwidth]{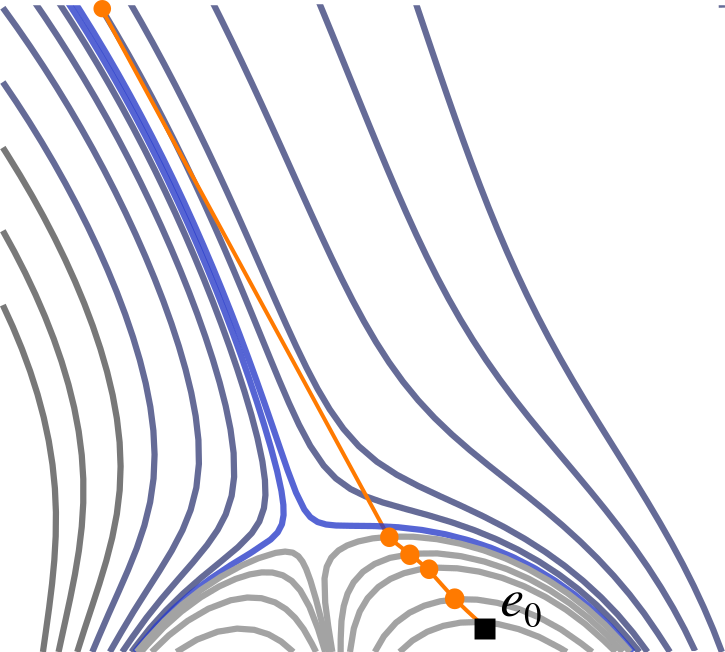}
    \caption{Panel (a): 2-dimensional sketch representing the `topological' path (orange solid line). Magnetic closed line apexes are denoted by dark edged rhomboids and the starting point with a red cross. Panel (b): Sketch of the 3D scenario. Only magnetic field lines crossing the orange cylinders are considered. The apex of each magnetic closed line is denoted by rhomboids as in panel (a) and edge rhomboids represent the minimum of the apexes inside the cylinder. Panel (c): Example of the `topological' path (orange line connecting the orange dots) calculated in a 2D example from the initial position $e_0$. Grey lines indicate closed field lines in a pseudostreamer topology (and helmet streamer to the left) and the blue lines are open field lines.
}
    \label{fig:top-sketch}
\end{figure}

\begin{figure*}[t]
    \centering
    \includegraphics[width=0.93\textwidth]{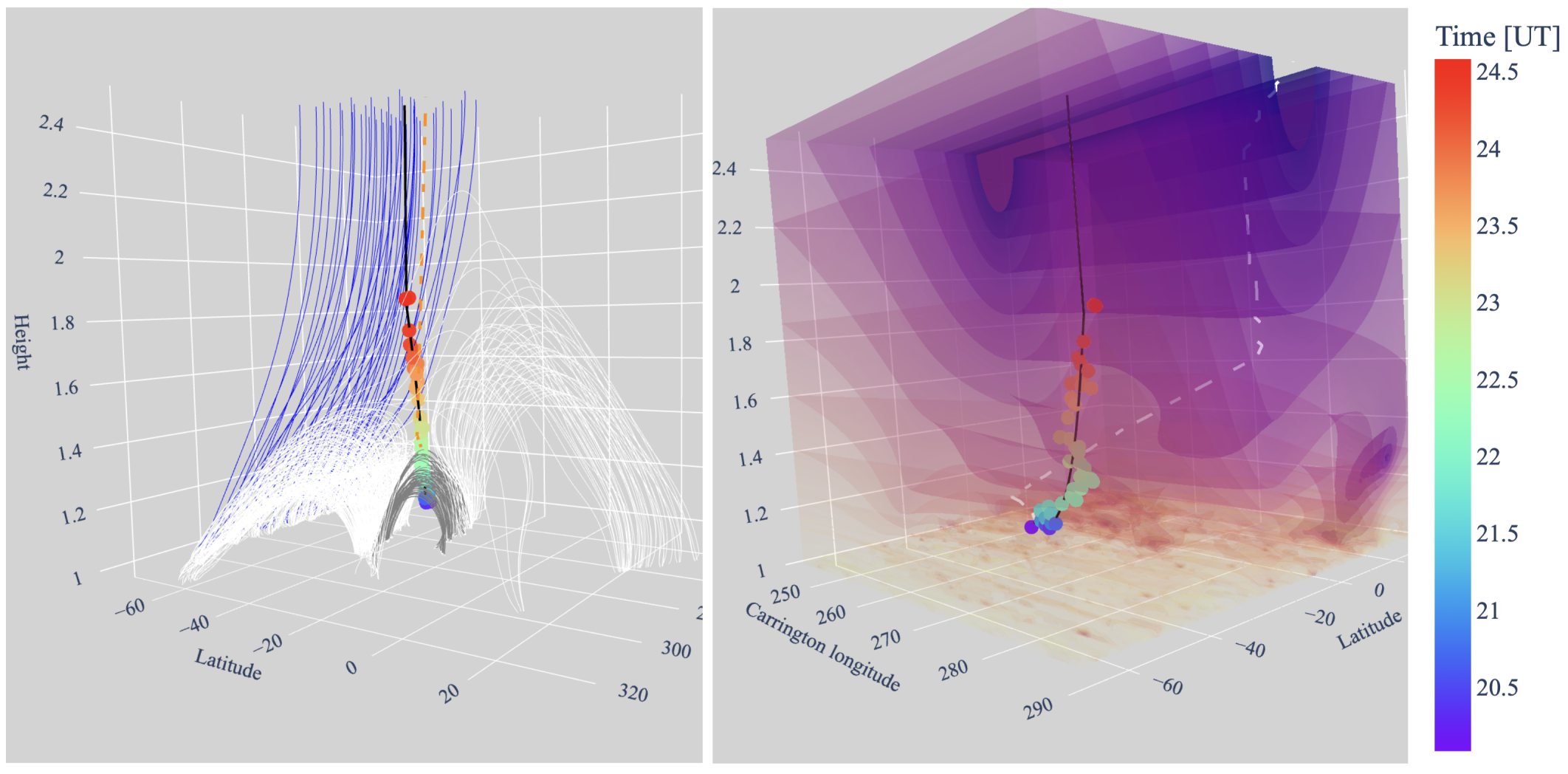}
    
    \caption{Magnetic ambient of the 2021-03-21 event. The left panel shows the topology of the field lines (white and grey lines correspond to closed field lines and blue lines to negative open lines) and the topological path (orange dashed line). The grey lines correspond to the arcade overlying the source region. The colored dots show the triangulated trajectory and the black line shows the eruptive path of the event, as Figure~\ref{fig:track_fit}. Right panel show the isocontours for the magnitude of the magnetic field being the yellowish more strong and decaying toward the purple contours. The white dashed line indicates the gradient path.}
    \label{fig:grad-top}
\end{figure*}
Considering the capacity of ambient magnetic field lines to guide and deflect solar eruptions, we introduce an alternative deflection path based on the topology of those field lines. 
We define this path as a sequence of points connecting the cusp of the field lines that enclose the source region of the eruption. Figure~\ref{fig:top-sketch}a shows a 2-dimensional view of the `topological' path. In this simplified scenario, the ``topological'' path is traced as the curve that connects the position of the maximum radius of each closed and open ($R_\text{max}=2.5\,R_\sun$) field line overlying the initial position. The field lines under consideration can be selected as those that cross the vertical line at $x_0$, with the initial position at $(x_0,y_0)$. In the 3D scenario (see Figure~\ref{fig:top-sketch}b), the field lines are not necessarily restricted to a single plane and can vary in two tangential coordinates, resulting in a non-planar trajectory. Instead of considering a magnetic field line located above the initial position, we consider the magnetic lines that cross a cylinder situated on top of it (Figure~\ref{fig:top-sketch}b). Because of the additional degree of freedom in the 3D case, it is necessary to update the position from which we calculate the cusp of the next field line. The new position corresponds to the apex of the magnetic field line that, out of all lines intersecting the cylinder on top of the initial position, presents the minimum radius. This enables a deflection toward the growing loops. Figure~\ref{fig:top-sketch}b shows an scheme of the volume considered (orange cylinder) and the point selected for the `topological' path (dark diamond).  Figure~\ref{fig:top-sketch}c shows an example of the `topological' path calculated from the initial position $e_0$. The final `topological' path will depend on the following selection criteria such as: the size of the area over the initial position, the density of lines, and the height of the closed field lines. 

The total computation of the paths, with the specific algorithm, for each event can be seen at \url{https://github.com/asahade/Events}.

\section{Results} \label{sec:results}

For each event on Table~\ref{tbl-events} we compute the `topological' and `gradient' paths as explained above. 
Figure~\ref{fig:grad-top} shows a detail of the ambient magnetic field lines (left panel) and isocontours of magnitude $B$ of the magnetic field (right panel) for the example event (2021-03-21). The colored dots mark the triangulated position and the solid black line marks the eruptive path (as described in Section~\ref{ss:eruptiveP}). In the left panel, we see that the filament lies underneath a low-lying arcade (gray lines). The ambient magnetic field consists of a system of three arcades and a coronal hole of negative polarity to the south (left). The white arcades (to the north and south of the source region) have opposite polarities and converge to a null-point corridor over the source region arcade (gray). Not displayed in the figure, open field lines of positive polarity converge towards the southern coronal hole, into the HCS. In this example, the `topological' path (orange dashed line), which tells us the trajectory of an eruption deviated only by the ``channeling'' of the field lines, departs from the source region towards the growing cusps of the gray arcade, passes throughout the null corridor separating the arcade fluxes, and finally follows the open field lines orientation of the southern CH. In the right panel of Figure~\ref{fig:grad-top} we notice the presence of null points (in dark purple) and the null corridor mentioned above by the lower height of the isocontours of $B$. The lowest magnetic energy region (darkest purple) is located at the HCS. From the source region location, the `gradient' path (which describes the trajectory of an eruption only deviates towards the minimal energy regions) goes to a local minima to the east and then towards the null corridor to after ends in the HCS.
The code and data to produce the corresponding figure for each event is provided in the repository (\url{https://github.com/asahade/Events}). Clearly, the `topological' path better represents the observed path of the 2021-03-21 eruption than the `gradient' path.

\begin{figure*}
    \centering
    \includegraphics[width=0.96\textwidth]{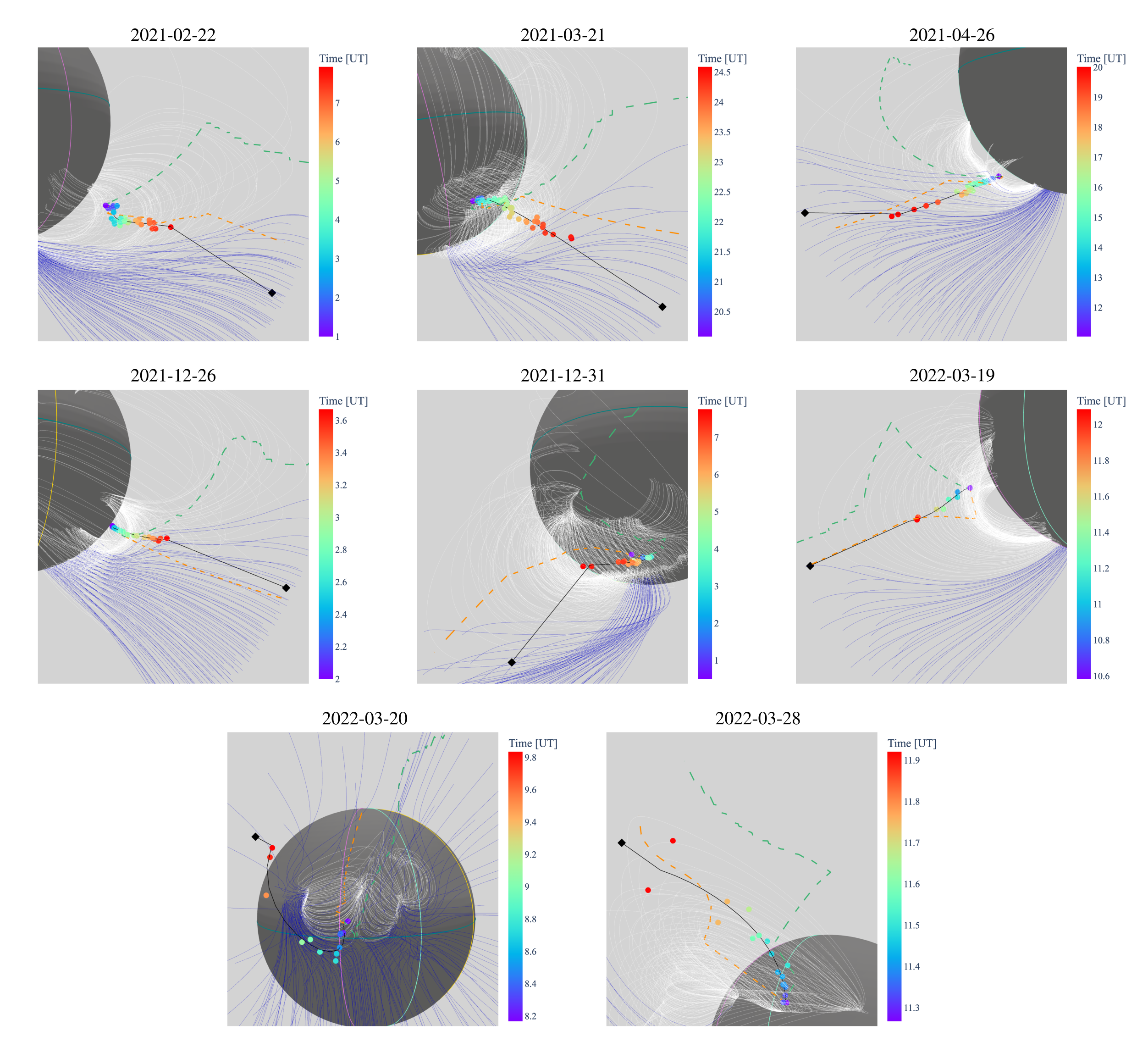}
    \caption{Projections of the `eruptive' (black solid line), `topological' (orange dashed line) and `gradient path' (green dashed line) for the studied events. Color dots idem Figure~\ref{fig:all-events}. The white lines correspond to closed field lines and blue lines to open lines. Over the sphere of $1\,R_\sun$, we plot the equator (teal line), STA meridian (pink line), SolO meridian (gold line), and SDO meridian (light green line) for reader reference.}
    \label{fig:3d_paths}
\end{figure*}

\begin{figure}
    \centering
    \includegraphics[width=0.47\textwidth]{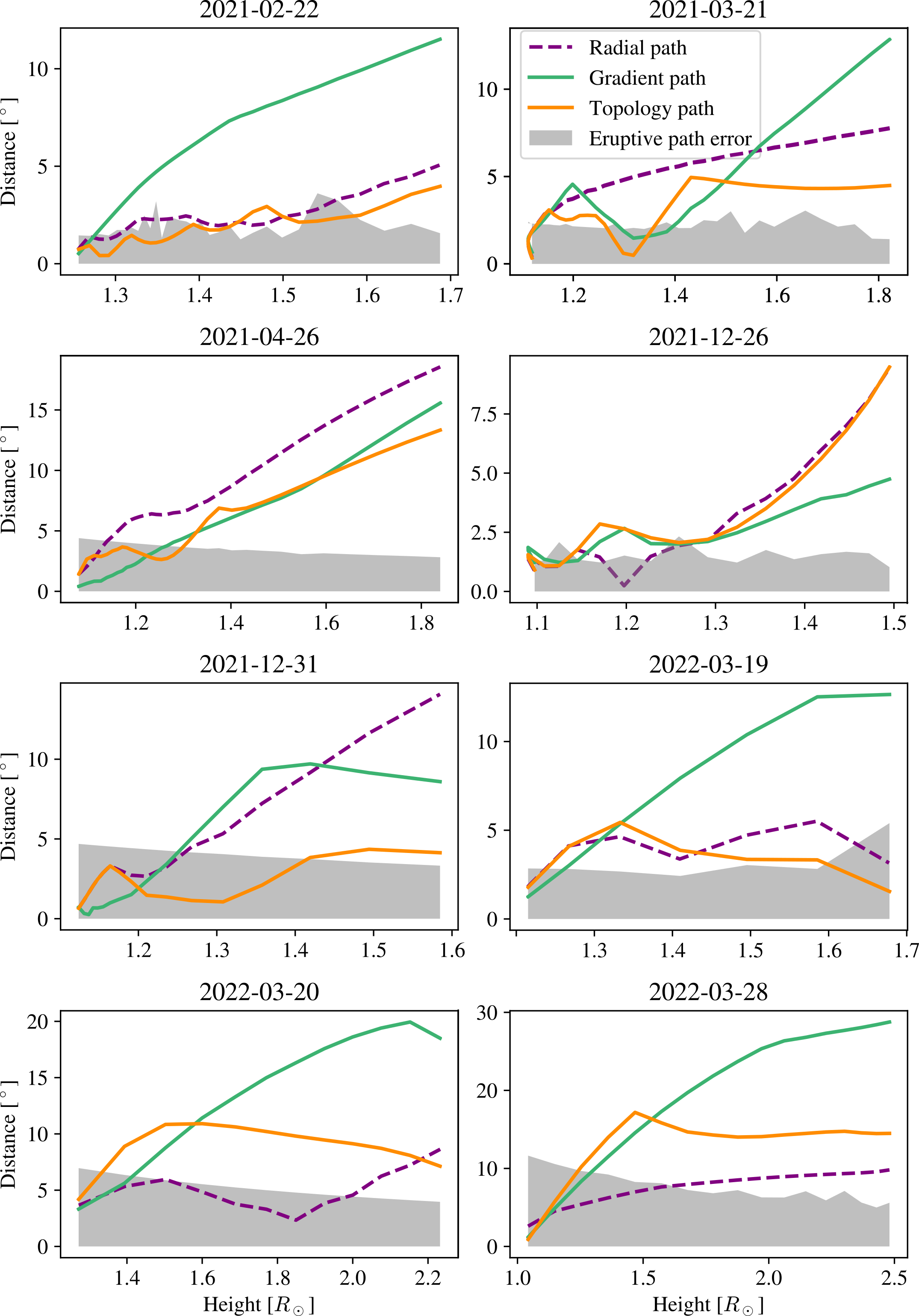} \\
    \includegraphics[width=0.47\textwidth]{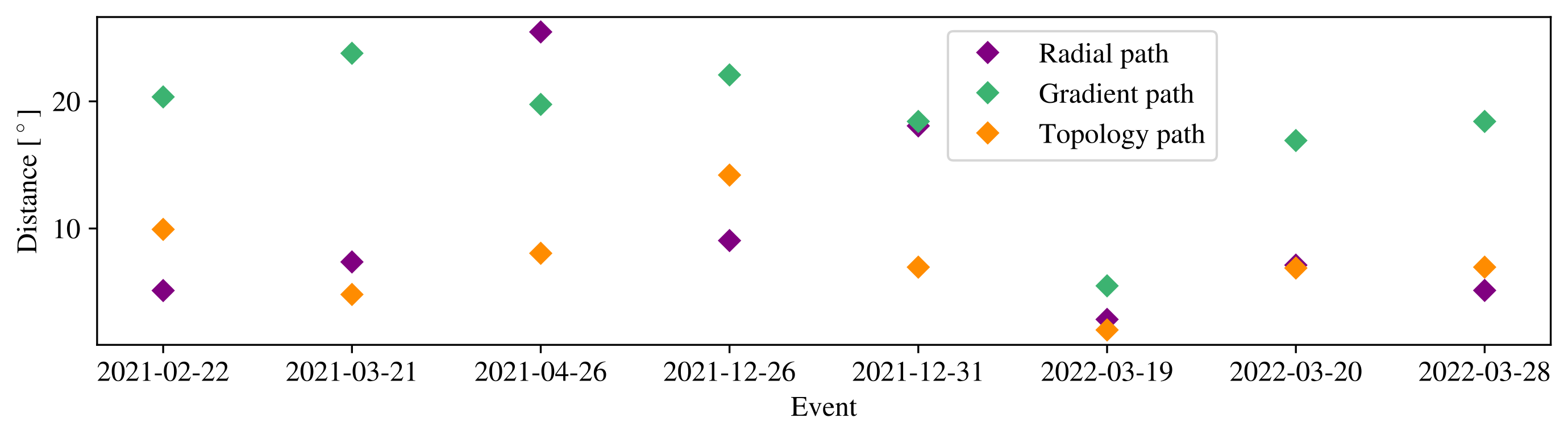}
    \caption{Distance between the `gradient' path (green line), and the `topological' path (orange line) respect to the eruptive path, as a function of the height (without including the GCS reconstruction). Purple dashed line denote the difference between the radial path from the source region and the actual eruptive path. The gray shaded area indicates the error of the eruptive path. \textit{Lower panel:} Angular separation to the eruptive path for all the events at $2.5\,R_\sun$.}
    \label{fig:path_dist}
\end{figure}

Figure~\ref{fig:3d_paths} shows the resulting paths for all events in Table~\ref{tbl-events}. Each panel displays a view of the tracked evolution (colored dots and black diamond as in Figure~\ref{fig:all-events}), together with the eruptive path (black solid line). Some relevant closed (white) and open (blue) field lines are drawn in the area of the eruption. Finally, the green dashed lines indicate the `gradient' paths and the dashed orange lines indicate the `topological' paths. Depending on the view, some reference lines can be seen over the sphere: equator (teal line), STA meridian (pink line), SolO meridian (gold line), and SDO meridian (light green line). 

We note that the `gradient' paths deflect towards the global minima, eventually ending at some part of the HCS. The gradient paths after $\sim 1.5\,R_\sun$ show a great discrepancy with the eruptive paths for all the cases. On the other hand, the `topological' paths deflect towards the top of overlying arcades, ending within the overlying open field lines. This not necessarily coincide with the center of the HCS, but some place near it. In general, the `topological' paths remain closer to the eruptive paths during the whole evolution.

To have a better notion of which agent, the gradient or the topology, has more influence in the deflection of the trajectory of solar eruptions, we quantify the distances between the ``gradient'' and ``topological'' paths with the eruptive one. This also gives us an idea of the performance of each of these paths to estimate the trajectory from its source region location.

Figure~\ref{fig:path_dist} shows the difference between the observed trajectory and the `gradient' (green line) and `topological' (orange line) paths. The upper panels zoom in the low corona, up to the last tracked position of the filament, and the lower panel summarizes the angular distance from the eruptive path, at $2.5\,R_\sun$, of each event. The purple dashed line represents the difference between the source radial and the actual eruptive path. This `radial' path indicates the degree of deflection and provides a sense on how far from radial propagation lies the observed trajectory. The gray shaded area indicates the standard deviation associated with the eruptive path. In all cases, the `gradient' path veers away from the eruptive path above $\sim 1.5\,R_\sun$, once the large structures (helmet streamers, coronal holes) of the magnetic field dominate. On the other hand, the `topological' path remains close to the observed path.
Compared against radial propagation assumption, the `gradient' path gives a better proxy for only the 2021-04-26 event, which exhibited the greatest deflection of the analysed events. In contrast, the ``topological'' path better approximates the eruptive path in 4 cases, and provided a similar estimation than the radial path for other 2. For the 2021-02-22 and 2021-12-26 the `topological' path departed almost $10^\circ$ from the eruptive path.
The 2021-12-26 event (right second row of Figure~\ref{fig:path_dist}) is the only event where the `gradient' approach outperforms the ``topological'' one in the low corona. The eruptive filament was embedded in a multi-polar region overlaid by a coronal hole with a null corridor of magnetic energy (similar to the 2021-03-21 event, but in a double arcade system). The `topological' path could not capture the connection with that null corridor, as the `gradient' path did.

\subsection{Error estimation}
We tested the robustness of the `gradient' and `topological' path derivations under changes like: starting coordinates, PFSS grid spacing (for the gradient path), and field line density (for the topological path). If we change the starting position by few cells in any direction, the gradient path is very stable. This means that the variation in the final path does not dramatically vary under the expected error in the source region determination. For the topological path this is also true while the variations remain under a same topological region (inside a separatrix layer). Both paths were also shown to be very stable under changes in density (in the grid or field lines), in general, we estimated errors under $<1^\circ$ for both paths, which is lower than the error from the eruptive path calculation.

Other errors, like the ones introduced by the the quality of the synoptic magnetogram or the PFSS reconstruction method, are beyond the scope of this work.

\section{Discussion and Conclusions}\label{sec:conclusion}

It has been shown that the deflections (and evolution in general) of solar eruptive events are influenced by the ambient magnetic field \citep[ssee, e.g.,][, and references therein]{cecere2023}. 
Past research has pointed out two distinctive agents as physical triggers of these deviations: the gradient of the magnetic pressure and the topology of the field lines. In this work, we aim to address the question of which one of them could influences the evolution of 8 well-observed events. For this purpose, we propose two `magnetic paths' that could be followed by an eruption if, starting from the source region, the propagation is driven by solely the magnetic gradient or  the magnetic topology.

First, we developed a new tool to accurately track the evolution of the eruptive filaments and perform 3D reconstruction by leveraging the multi-viewpoint observations 
\citep[\texttt{scc\_measure3.pro},][]{scc_measure3}. Also, we modeled the CMEs related to those filaments, including when possible the SoloHI observations. 

Then, we analyzed the path of eruptions as predicted by the `gradient' path. We found that this path is generally very different than the actual trajectory, indicating that deflections (at $2.5\,R_\sun$) can be overestimated by as much as $\sim 20^\circ$. The `topological' path, on the other hand, better described the eruptive path of the studied events. This results suggests, rather strongly, that the topology of the surrounding magnetic field may be the dominant driver for the deflection of  eruptions in the low corona. Of course, the two quantities are interconnected, as we are considering a potential magnetic field, but the topology considers not only the strength of the magnetic field but also its connectivity and `shape.' Clearly, the latter are important factors to consider in the evolution of CMEs.

Therefore, we introduce a new method to interpret the very early path of an eruption. The topological and gradient method rely only on the source region location and 3D structure of the background coronal field. The CME measurements are not required. These properties makes the topological method a potentially powerful addition to forecasting infrastructure. We expect to extend the study presented in this work to more events to test the `topological' path as an estimator of CMEs deflection and then, its power to predict the direction of propagation only by knowing the source region of an event.

There are caveats to consider; the accuracy of the method relies on the accuracy of the field extrapolations, which, in turn, rely on accurate boundary conditions. The lack of vector field measurements above the photosphere and instantaneous global coverage (for regions beyond $60^\circ$) reduces the applicability of this technique (and of the 'gradient' technique as well) to a limited range of source region locations. In addition, the dynamic evolution of the ambient field configuration during eruption will also reduce the reliability of both techniques, especially in the cases of multiple eruptions in close sequence and intense activity throughout the source region. Lastly, we should recall that these methods only consider the ambient effects on the event propagation. They do not consider internal evolution or interactions among multiple eruptions. Despite those caveats, our analysis represents an improvement from the default assumption of a radial evolution.

The promising results from this work indicate that stereoscopic missions (both magnetic field and coronal imaging) may be key for obtaining reliable predictions of CME trajectories \textit{before\/} event occurrence. The need for such observations has been discussed in the context of space weather research \citep{Schrijver2015} and gap analysis \citep{Vourlidasetal_2023}. 

\begin{acknowledgements}
    AS was supported by an appointment to the NASA Postdoctoral Program at the NASA Goddard Space Flight Center, administered by Oak Ridge Associated Universities under contract with NASA. AV was supported by NASA grant  80NSSC21K1860. CM was supported by an appointment to the Solar Orbiter Heliospheric Imager (SoloHI). SoloHI instrument was designed, built, and is now operated by the US Naval Research Laboratory with the support of the NASA Heliophysics Division, Solar Orbiter Collaboration Office under DPR NNG09EK11I.
    We acknowledge the use of Solo/EUI, SoloHI, SDO/AIA, SOHO/LASCO, and STEREO/EUVI, COR2 data.
 \end{acknowledgements}

\appendix

We present here two examples of the utility of a third point of view for determining the 3D coordinates of a feature by the tie-pointing technique. A more extensive analysis of the limitations and errors induced by the usual two-viewpoint tie-pointing can be found in \citet[e.g., ][]{Inhester2006,Mierla2010}.

\begin{figure}    
   \centerline{\includegraphics[width=0.49\textwidth]{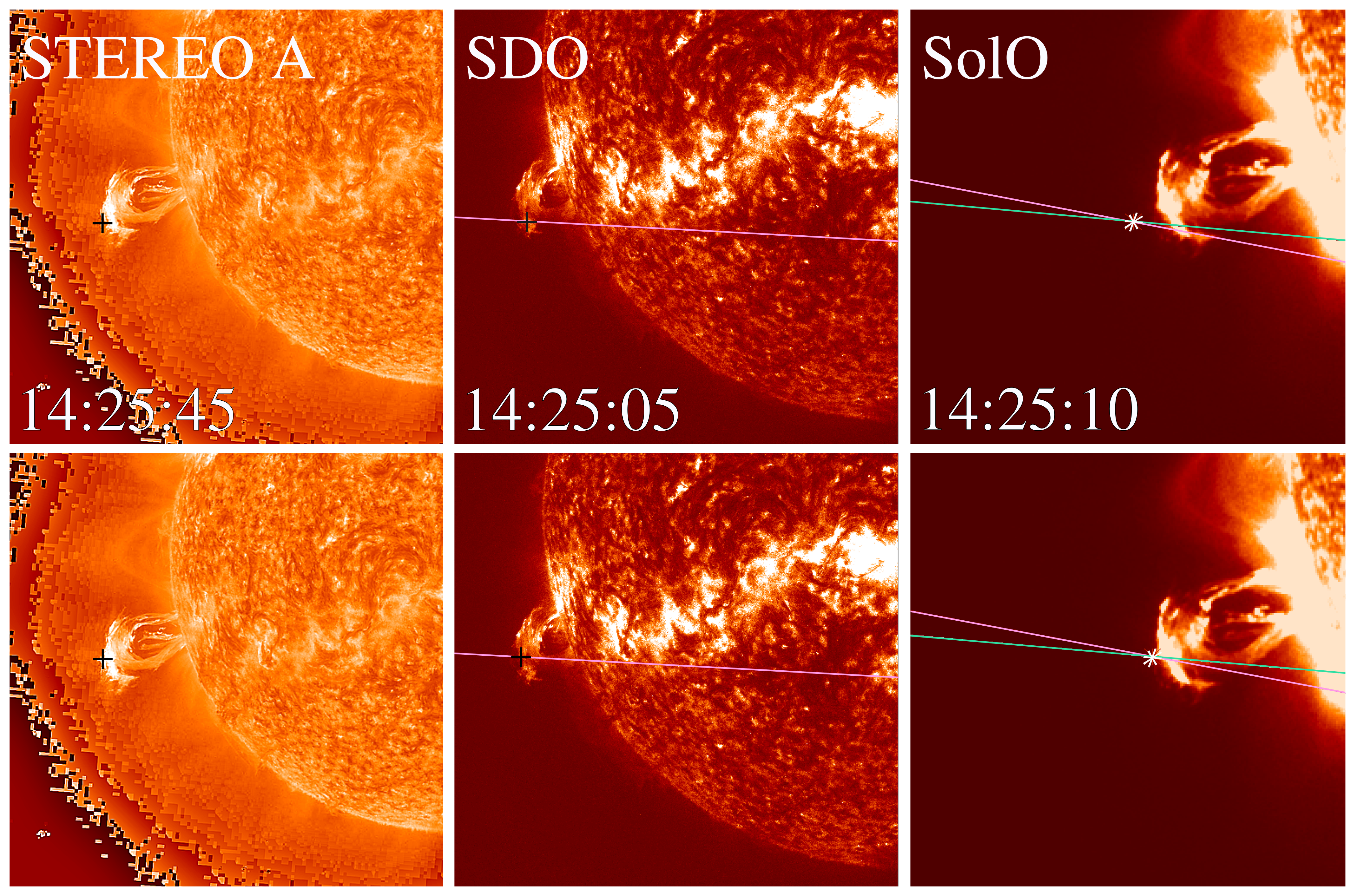}
             }
 \caption{Tie-pointing with the \texttt{scc$\_$measure3} routine applied to STEREO A (left panels), SDO (central panels), and SolO (right panels) EUV images. A pixel selected over the prominence projection of the STEREO A image (black cross), produce a epipolar line (pink) over the other two spacecraft images. In the central panels a second pixel is selected over that line (black cross in pink line), producing a second epipolar line (green) over the SolO image, that crosses at a unique point (white asterisk). Upper and lower panels show two different selections and consequent projection of the 3D coordinate over the SolO image. } 
\label{fig:scc3_a}
\end{figure}

The tie-pointing technique determines the position of a feature in 3D by identifying the same feature appearing in the two A and B images. The projections of a feature will lie along corresponding epipolar lines, which reduces the triangulation of a point to a one-dimensional search. This means that when we select a pixel in the image taken by spacecraft A, its epipolar line is projected over the image taken by spacecraft B, then if we identify the same characteristic over that line we will successfully determine its 3D coordinates. 
However, the reconstruction can be ambiguous if we are not able to determine correspondences between the projections. Two common sources of ambiguity are the extended nature of the element that we want to locate in 3D space and the transparency of the coronal plasma.   With the routine \texttt{scc$\_$measure3} we include a third image to constrain and validate the 3D reconstruction. Figure~\ref{fig:scc3_a} shows an example of the routine for an eruptive filament observed simultaneously on the limb of STEREO A (left panels), SDO (central panels) and SolO (right panels) at 14:25 UT. It is worth noting that the observation date of SolO was corrected to the time observed at 1 AU, since the spacecraft observed the event earlier than the other missions because it was closer to the Sun ($d_\sun=0.9$ AU). On that date the separation between SDO and STEREO A was $\gamma_a=-13^\circ$, and between SDO and SolO was $\gamma_b=-31^\circ$. In the upper panels a pixel (black cross) is selected in the STEREO A image producing a epipolar line (pink line) overlaid on the SDO and SolO images, then over that line another pixel is selected in the SDO projection tracing that epipolar line (teal) over the SolO image; the asterisk indicates the projection of the triangulated position. Clearly, the reconstruction has an offset from the real position of the feature as can be seen in the right upper panel, in which the asterisk and the filament are not at the same place. As the asterisk does not match the position of the prominence in SolO projection we can discard the reconstruction and repeat the process as shown in the lower panels.

\begin{figure}    
   \centerline{\includegraphics[width=0.49\textwidth]{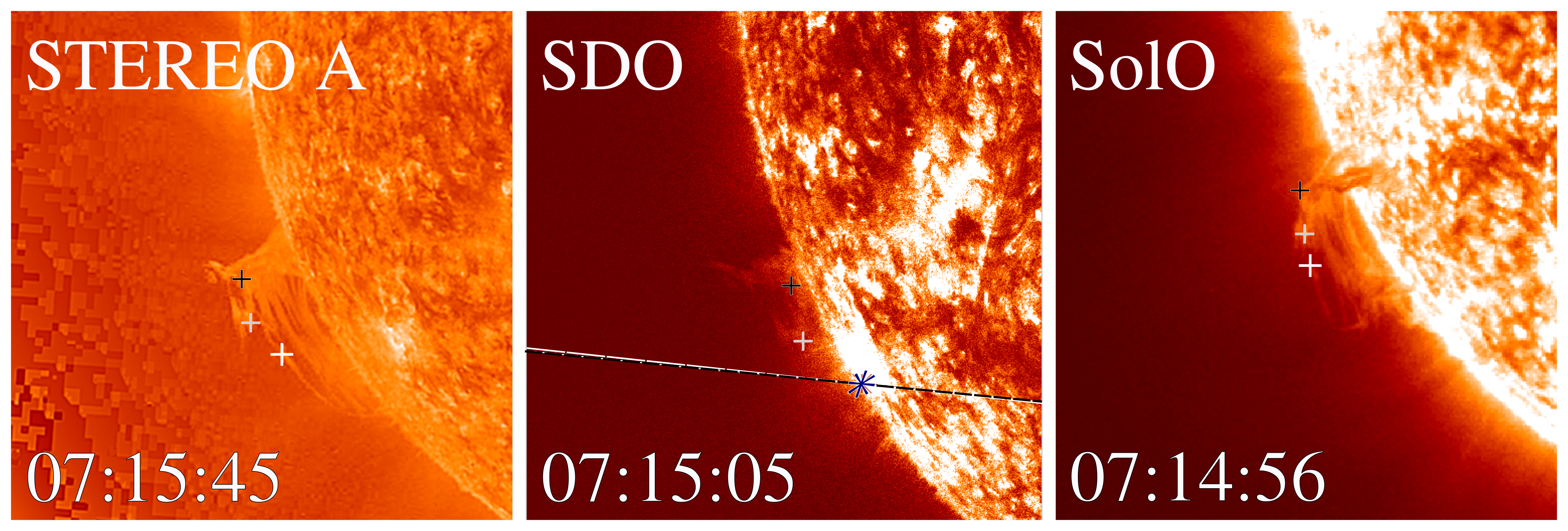}
             }
 \caption{Tie-pointing with the \texttt{scc$\_$measure3} routine applied to STEREO A (left panel), SDO (central panel) and SolO (right panel) images. Different parts of an extended prominence are triangulated with crosses in black, grey and white, respectively. The white crosses in the STEREO A and SolO images correspond to the blue asterisk on the SDO image. \textbf{The epipolar projections are very close together and to better distinguish them the epipolar line of STEREO A is plotted in dashed white and SolO epipolar line in dashed black.} } 
\label{fig:scc3_b}
\end{figure}

Another advantage of including a third view-point is the triangulation of an extended element, such as a prominence. The routine allows us to select the first and second positions from any images, leaving the third one as the control image. While a pair of images can help to identify parts of the filament, another pair can help to determine the position of another segment. Figure~\ref{fig:scc3_b} shows an example where a prominence over the limb is triangulated. The positions of the black and gray cross are easier to establish from the STEREO A and SDO images, since the projections of the features in both images are alike. However, for the white crosses overlaid on the STEREO A and SolO images, the corresponding position in SDO (blue asterisk) is very difficult to infer. Conversely, the white and grey crosses position are difficult to infer from the SolO projection. 

The routine is publicly available at \url{https://github.com/asahade/IDL}; if used please cite \cite{scc_measure3}.
\bibliography{biblio}{}
\bibliographystyle{aasjournal}

\end{document}